\documentclass[aps,showpacs,twocolumn]{revtex4}
\usepackage{graphicx,amssymb,amsbsy,color}
\begin{document}

\title{Spin Hall current and two-dimensional magnetic monopole in a Corbino disk}

\author{Ping Lou$^{1,2}$ and  Tao Xiang$^{1}$}

\affiliation{$^{1}$ Institute of Theoretical Physics and
Interdisciplinary Center of Theoretical Studies, Chinese Academy
of Sciences, P. O. Box 2735, Beijing 100080, China}

\affiliation{$^{2}$ Department of Physics, Anhui University, Hefei
230039, Anhui, China}

\date{\today}

\begin{abstract}
A spin Hall current can circulate in a Corbino disk with
spin-orbit interactions in the presence of a radial charge
current, without encountering the sample edge problem. We have
suggested several experiments to directly detect and manipulate
the spin Hall current in this system. A fictitious two-dimensional
magnetic monopole can be created in a Corbino disk with the Rashba
interaction by a circular charge current induced either by a spin
Hall current in the presence of a ferromagnetic junction or by a
perpendicular external magnetic field.
\end{abstract}

%\pacs{72.25.Fe, 73.21.Fg, 73.63.Hs, 78.67.De}
\maketitle

Recently great attention has been paid to the generation and
manipulation of spin currents in
semiconductors.\cite{Murakami,Sinova,Hirsch,Zhang,Dyakonov,Rashba2,Schliemann,Shen,Mishchenko,Sheng,Zutic}
A fundamentally interesting and technically promising proposal is
that a spin current can be created by the spin Hall
effect,\cite{Murakami,Sinova,Hirsch} arising purely from the
relativistic spin-orbit interactions. This can avoid the loss of
the coherence of polarized spin current injected from a
ferromagnetic metal at the interface due to the conductance
mismatch.\cite{Zutic} When a metallic sample is exposed to an
external magnetic field, the Lorentz force acting on the current
carriers gives rise to a transverse voltage. Similar to the charge
accumulation in the conventional Hall effect, spin accumulation is
expected at the sample edges in the spin Hall effect. This spin
accumulation has been observed recently by Kato {\it et
al.}\cite{Kato2} and by Wunderlich {\it et al.}\cite{Wunderlich}.
The internal magnetic field \cite{Kato} and the spin polarization
\cite{Kato1} induced by the spin-orbit coupling were also observed
in bulk semiconductors. However, as the spin current cannot
circulate within the devices measured, the spin Hall current
itself has not been observed directly. It is also controversial
whether the observed spin accumulation results from intrinsic or
extrinsic spin Hall effects.\cite{Kato2,Bernevig2}

In this paper, we propose to use a radial current to generate a
circulating spin Hall current in a Corbino disk. This would allow
the spin Hall current to be directly detected and manipulated.
Furthermore, it will be shown that a fictitious two-dimensional
magnetic monopole can be created in this Corbino disk system,
resulting from the interplay between the spin-orbital interaction
and a transverse charge current induced either by a spin Hall
current in the presence of a ferromagnetic junction or by a
perpendicular external magnetic field. Topologically, a Corbino
disk is equivalent to a cylindric system if we take the inner and
outer boundaries of the Corbino disk as the two ends of the
cylinder. Thus with proper changes of boundary conditions most of
the results given below can be also applied to a cylindric system.

\begin{figure}[[htbp]
\begin{center}
\includegraphics[width=7cm,trim=0 250 0 250,clip=,angle=0]{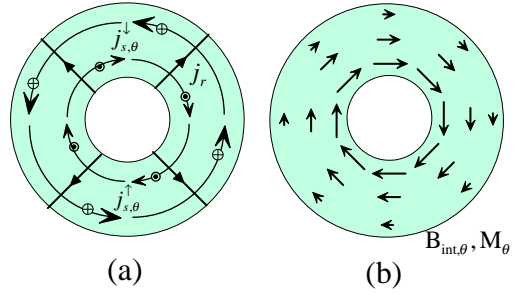}
\caption{(a) Corbino disk with an applied radial charge current
$j_r$ and an induced circular spin Hall current polarized
perpendicular to the disk plane between the two contacts. The spin
Hall current can be taken as a combination of a current of up-spin
electrons in the clockwise direction $j_{s,\theta}^\uparrow$ and
that of down-spin electrons in the opposite direction
$j_{s,\theta}^\downarrow$. (b) The vortex-like internal magnetic
field $B_{\text{int},\theta}$ or spin polarization $M_{\theta}$
generated by the applied charge current in a Rashba system. }
\label{fig1}
\end{center}
\end{figure}

A Corbino disk is an annular region of conducting materials
surrounding a metallic contact and surrounded by a second metallic
contact (Fig. \ref{fig1}a). It has played an important role in the
thought experiments that first clarified the nature of the integer
quantum Hall effect\cite{Laughlin,Halperin} If a radial current
$I$ is applied between the two contacts, it will generate an
electric field along the direction of current
\begin{equation}
E_r (r) = \frac{I}{2\pi r w \sigma_0}, \qquad (a\leq r \leq b),
\end{equation}
where $\sigma_0$ is the longitudinal conductance at zero magnetic
field, $w$ is the thickness, $a$ and $b$ are the inner and outer
radii of the Corbino disk, respectively. The corresponding voltage
drop between the two contacts is
\begin{equation}
V_{0} =  \frac{I}{2\pi  w \sigma_0}\ln \frac{b}{a}.
\label{eq:voltage}
\end{equation}

In a spin Hall Corbino disk, a radial electric field will generate
a transverse spin current with spins polarized perpendicular to
the disk plane. This spin current can circulate persistently
within the disk, powered by the radial current $I$. The circular
spin current density $j_{s,\theta}$ is proportional to $E_r$.
Assuming $\sigma _{s,\theta }$ to be the spin Hall conductance,
induced either by intrinsic or by extrinsic Hall effects, then
$j_{s,\theta}$ is given by
\begin{equation}
j_{s,\theta } (r) = \sigma_{s,\theta} E_r(r) =
\frac{\sigma_{s,\theta } I }{ 2\pi w r\sigma_0}.
\label{eq:spincurr1}
\end{equation}
The corresponding total spin current is
\begin{equation}
I_{s} = d\int_{a}^{b}\text{d} r j_{s,\theta } (r) =
\frac{\sigma_{s,\theta } I}{2\pi \sigma _{0}}\ln  \frac{ b}{a}.
\end{equation}
This spin current can be taken as a combination of a current of
clockwise-circulated up-spin electrons and a current of
contraclockwise-circulated down-spin electrons. It results in a
flow of spin angular momentum without net charge current.

In the ordinary Hall effect, a transverse force perpendicular to
an electric field is provided by the Lorentz force from a magnetic
field. In a spin Hall system, this transverse force arises from
the relativistic spin-orbit interaction. The spin-orbit coupling
does not depend on a magnetic field. It opens a new route to the
manipulation of spins with non-magnetic semiconductors without
external magnetic fields.

In a two-dimensional electron gas of semiconductors, the
spin-orbit interaction arising from the asymmetry in the growth
direction ($z$) has a standard Rashba form.\cite{Rashba} Recently,
this kind of spin-orbit interaction has attracted special
attention, in part due to the work of Sinova {\it et al.} who
showed that an intrinsic and universal spin Hall current can be
induced by this interaction.\cite{Sinova} The Hamiltonian of an
electron gas with the Rashba spin-orbit coupling is defined by
\begin{equation}
\hat{H} =\sum_{k} \left( \frac{\hbar ^{2}k^{2}}{2m}- \alpha _{R}
\mathbf{e}_{z} \cdot \mathbf{k} \times \mathbf{\sigma} \right),
\label{eq:Rashba}
\end{equation}
where $\mathbf{k}$ is the wavevector and $\mathbf{\sigma}$ is the
Pauli matrix. The Rashba interaction couples the orbital motion of
electrons with their spins. Effectively, it can be taken as a
Zeeman coupling of a spin subjected to the interaction of an
internal magnetic field. Thus the above Hamiltonian can be also
expressed as
\begin{equation}
\hat{H} = \sum_k\left[ \frac{\hbar ^{2}k^{2}}{2m}-\frac{1}{2}g\mu
_{B}\mathbf{B}_{\text{int}}\mathbf{(k} )\cdot \mathbf{\sigma }
\right] ,
\end{equation}
where  $\mu _{B}$ is the Bohr magneton, $g$ is the gyromagnetic
ratio, and
\begin{equation}
\mathbf{B}_{\text{int}}\mathbf{(k})= \frac{2\alpha _{R}}{g\mu
_{B}} \mathbf{e}_{z} \times \mathbf{k}
\end{equation}
is the effective internal magnetic field.

Without an external electric field, the internal field
$\mathbf{B}_{\text{int}}$ in real space is zero since the average
of the momentum $\mathbf{k}$ is zero over the Fermi sea. However,
by applying an electric field to the system, the average of
$\mathbf{k}$ becomes finite in the direction opposite to the
applied field. In this case a finite magnetic field perpendicular
to both the applied electric field and the disk plane will be
generated by the Rashba spin-orbital coupling. Thus for the system
considered here, a radial electric current will induce an internal
magnetic field in the radian direction. Under the semiclassical
approximation, it can be readily shown that this internal magnetic
field is given by $\mathbf{B}_{\text{int}}(\mathbf{r}) =
B_{\text{int}, \theta}(r) \mathbf{e}_{\theta}$ and
\begin{equation}
B_{\text{int}, \theta}(r) = -\frac{A}{2 \pi r} , \label{eq:field1}
\end{equation}
where $A=(2e\tau \alpha _{R}I) / (\hbar g\mu _{B} w \sigma_0 ) $
and $\tau$ is the relaxation time of electrons. As in the usual
electron gas system, this internal magnetic field will polarize
the electrons and lead to a finite spin magnetization at each
point parallel to $\mathbf{B}_{\text{int}}(r)$. From the spin
susceptibility of a two-dimensional electron gas and Eq.
(\ref{eq:field1}), it can be shown that the spin magnetization
induced by the Rashba interaction is proportional to
$\mathbf{B}_{\text{int}}$ and given by
\begin{equation}
\mathbf{M}(\mathbf{r})= \frac12 g\mu_B \langle \mathbf{\sigma}
\rangle = -\frac{e\alpha _{R}m\tau g\mu _{B}}{4\pi \hbar
^{3}}E_{r} \mathbf{e}_{\theta} . \label{eq:mag1}
\end{equation}

Fig. (\ref{fig1}b) shows the spatial distribution of
$\mathbf{B}_{\text{int}}$ or equivalently $\mathbf{M}$. The
spatial dependence of $\mathbf{B}_{\text{int}}(\mathbf{r})$ is
similar to the magnetic field generated by a line current along
the z-axis at $r=0$. But $\mathbf{B}_{\text{int}}$ exists only
inside the sample. This kind of internal magnetic field as well as
the spin polarization induced by the spin-orbit coupling were
observed in bulk semiconductors by the time and spatially resolved
Faraday or Kerr rotation.\cite{Kato,Kato1,Kato2}

The Rashba coefficient $\alpha_R$ can be tuned over a wide range
by a vertical electric field or a strain pressure.\cite{Crooker}
In bulk semiconductors without strain, the internal magnetic field
produced by an applied electric current and the Rashba interaction
is generally less than $10^{-6} \text{T}$ and difficult to be
detected experimentally\cite{Kato2}. However, by adding pressure
or strain to the sample, the Rashba coupling $\alpha_R$ is
dramatically enhanced and the internal magnetic field can be as
high as $10^{-3} \text{T}$.\cite{Kato2} Thus in a Corbino disk of
strained semiconductor, the predicted spatial distribution of the
internal field (\ref{eq:field1}) or spin polarization
(\ref{eq:mag1}) can be easily detected by the Kerr rotation
experiment. The induced spin polarization can be also probed
directly by scanning SQUID (superconducting quantum interference
device) measurements.

In a pure Rashba system, the longitudinal conductivity $\sigma_0$
is given by
\begin{equation}
\sigma _{0}=\frac{e^{2}n\tau }{m}-\frac{me^{2}\tau }{2\pi \hbar
^{4}} \alpha _{R} ^{2}, \label{eq:sigma}
\end{equation}
where $n$ is the electron concentration. The first term is the the
usual Drude term. The second term is the correction to the Drude
conductivity from the Rashba spin-orbit interaction. In real
materials, there are always impurities. The spin-orbit interaction
resulting from the gradient of the impurity scattering potential
can affect both the spin Hall effect and the charge dynamics of
the system. As shown by Hirsch\cite{Hirsch}, the skew scattering
of a spin current by impurity spin-orbit interaction can lead to a
transverse charge imbalance, giving rise to a Hall voltage. This
will add an impurity dependent term to Eq. (\ref{eq:voltage}) and
change the voltage between the two contacts from $V_0$ to
\begin{eqnarray}
V=\left( 1- \frac{4\pi R_{s}g \mu _{B}e n \sigma _{s,\theta
}}{\hbar} \right)^{-1} V_{0}, \label{eq:cvoltage}
\end{eqnarray}
where $R_{s}$ is the anomalous Hall coefficient. If the anomalous
Hall effect is induced purely by impurity scattering in this
paramagnetic system, then $R_s$ is expected to be proportional to
the impurity concentration. Experimental measurement of this
impurity-induced correction to the voltage fall between the inner
and outer contacts will provide a direct detection of the
circulated spin current. Furthermore, by varying the Rashba
coupling, for example, by applying a vertical electric field or a
strain pressure, one can also determine whether the spin Hall
current is mainly generated by the intrinsic spin-orbit coupling
or by the extrinsic impurity scattering from this kind of
measurements.

\begin{figure}[[htbp]
\begin{center}
\includegraphics[width=8.5cm,trim=45 300 30 300,clip=,angle=0]{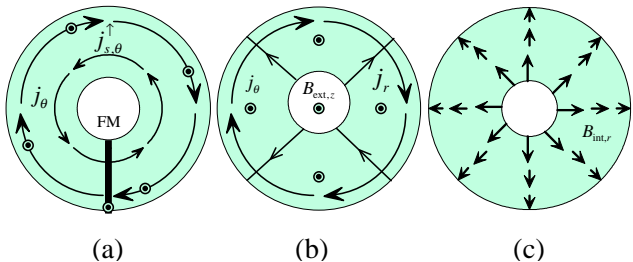}
\caption{(a) Corbino disk with a narrow ferromagnetic junction
polarized along the $z$ direction. The width of the junction is
assumed to be smaller than the spin diffusion length. An intrinsic
charge Hall current can be generated by the spin Hall effect if
only up-spin electrons can tunnel through the junction. (b)
Corbino disk in a perpendicular external magnetic field. Besides
the applied current $j_{r}$ between the inner and outer contacts,
there is also a transverse Hall current $j_{\theta}$ maintained by
the Lorentz force. (c) The internal magnetic field induced by the
charge Hall current in the disk with Rashba spin-orbit
interactions. Effectively, this radial magnetic field can be taken
as the field created by a two-dimensional magnetic monopole at
$r=0$. But $B_{\text{int},r}$ is finite only inside the disk.}
\label{fig2}
\end{center}
\end{figure}

Direct measurement of the spin Hall current in the Corbino disk is
difficult since the spin current does not couple with
electromagnetic fields. However, one can measure the charge
current generated by the spin Hall effect if one can use a narrow
ferromagnetic (FM) junction to block the down-spin Hall current
and allow only the up-spin Hall current to flow inside the disk
(Fig. \ref{fig2}a). This requires that the FM junction is
magnetically polarized along the $z$-direction and the width of
the junction is smaller than the spin diffusion length. In this
case, only up spin electrons can tunnel through the junction and
down spin electrons will accumulate by the one side of the
junction facing the direction of the down-spin current.

The accumulation of down spin electrons will form effectively a
potential barrier to prevent the tunnelling of up spin electrons
through this accumulation area. Two limiting situations can happen
depending on the degree of accumulation. One is the strong
accumulation limit at which the up-spin Hall current is completely
blocked by the potential barrier. The other is the weak
accumulation limit at which the potential barrier is low and the
up-spin electrons can tunnel through this accumulation barrier to
form a persistently Hall current inside the disk.

The weak accumulation limit is physically more interesting. In
this limit, a Hall current of up spin electrons will flow in the
disk. This is a charge Hall current entirely induced by the spin
Hall effect, without any external magnetic field. This Hall
current density $j_\theta$ is expected to be proportional to the
spin Hall current $j_{s,\theta}$ determined by Eq.
(\ref{eq:spincurr1}), i.e
\begin{equation}
j_{\theta}(r)=-\frac{2e\delta}{\hbar}j_{s,\theta}(r),
\end{equation}
where the dimensionless coefficient $\delta$ is expected to be
smaller than 1/2. $j_\theta (r)$ can be measured experimentally.
An experimental confirmation of this transverse charge current
without external magnetic fields would be a direct and unambiguous
proof of the spin Hall effect.

As for $j_{r}$, the interplay between the Rashba interaction and
the Hall current $j_{\theta }$ will also induce an internal
magnetic field in the direction perpendicular to $j_\theta$ inside
the disk. Thus in this case the internal magnetic field contains
not only the $B_{\text{int},\theta } (r)$ term induced by $j_r$
(Eq. (\ref{eq:field1})), but also a radial term $B_{\text{int},r}$
induced by $j_\theta$, i.e. $\mathbf{B}_{\text{int}} (\mathbf{r})
= B_{\text{int},r} (r) \mathbf{e}_r + B_{\text{int},\theta} (r)
\mathbf{e}_\theta $. Within the semiclassical approximation,
$B_{\text{int},r}$ is given by
\begin{equation}
B_{\text{int},r}(r) = -\frac{2e\tau\alpha_R j_\theta (r)}{\hbar g
\mu_B\sigma_0} = \frac{e_g}{2\pi r}, \label{eq:field2}
\end{equation}
where $e_g = (2e A \delta\sigma_{s,\theta}^z ) / (\hbar
\sigma_0)$. This is a divergent magnetic field. It is like the
field generated by a two-dimensional magnetic monopole with a
``charge" proportional to $e_g$. A schematic illustration of the
distribution of $B_{\text{int},r}(r)$ is shown in Fig.
(\ref{fig2}c). If we set the inner radius $a\rightarrow 0$, then
it is straigtforward to show that the divergence of
$\mathbf{B}_{\text{int}}$ is given by
\begin{equation}
\nabla \cdot \mathbf{B}_{\text{int}} = e_g \delta (\mathbf{r}) .
\end{equation}
This is a fundamentally interesting result. It indicates that one
can create a fictitious magnetic monopole in a Corbino disk with
the Rashba spin-orbit interaction simply by applying a circular
current. This fictitious magnetic monopole results from the
interplay between the complex topology of the Corbino disk and the
spin-orbit interaction. We believe that a
multiple-fictitious-magnetic-monopole system can be similarly
created in a more complex topological system. This would provide
an opportunity to simulate experimentally the Maxwell equations in
the presence of magnetic monopoles.

One can also create a charge Hall current in the disk without the
FM junction, but by applying a vertical magnetic field
$B_{\text{ext},z}$, as shown in Fig. (\ref{fig2}b). Since the
Lorentz force produced by the external magnetic field acts
symmetrically on both the up and down spin electrons, now the
charge Hall current is not spin polarized. From the standard
theory of electromagnetic response, we know that the Hall current
density is given by
\begin{equation}
j_{\theta } (r) = \frac{ \omega _{c}\tau I}{2\pi w r},
\end{equation}
where $\omega _{c}= {eB_{\text{ext},z}}/{m}$ is the cyclotron
resonance frequency. As for the case shown in Fig. (\ref{fig2}a),
this circular current will also induce an internal magnetic field
along the radial direction. But now $B_{\text{int},r}$ becomes
\begin{equation}
B_{\text{int},r}(r) = \frac{ \omega _{c}\tau A}{ 2\pi r } .
\label{eq:field3}
\end{equation}
It can be also taken as a magnetic field generated by a fictitious
two-dimensional magnetic monopole. The ``charge" of the monopole
is now proportional to $\omega_c \tau A$.

In conclusion, a spin Hall current can circulate in a Corbino disk
in the presence of a radial charge current. Experimental
measurements of the voltage drop between the two contacts as
functions of $\alpha_R$ and $R_s$ would allow us to judge
unambiguously if the spin current is mainly produced by the
intrinsic Hall effect. $\alpha_R$ and $R_s$ can be tuned by
applying a vertical electric field or a strain pressure and by
altering the impurity concentration, respectively. The interplay
between the Rashba spin-orbital interaction and a circular charge
current, which can be created either by a spin Hall current in the
presence of a narrow ferromagnetic junction or by a perpendicular
external magnetic field, will induce a divergent internal magnetic
field. This divergent internal magnetic field can be effectively
taken as the field produced by a fictitious two-dimensional
magnetic monopole. Thus the Corbino disk with spin-orbit
interactions provides a real-space realization of the Dirac
magnetic monopole in a solid state system. It can be used as a
basic unit for further studies of Maxwell electrodynamics with
magnetic monopoles.

We are grateful to Z. B. Su and L. Yu for useful discussions. This
work was supported by the National Natural Science Foundation of
China.

\end{document}